\documentclass{elsart}
 \usepackage{graphics}
 \usepackage{graphicx}
 \usepackage{epsfig}

\newcommand{\lw}[1]{\smash{\lower2.0ex\hbox{#1}}}

\begin{document}
\begin{frontmatter}

\title{ On the Breit Equation }

\author[hk]{Hikoya Kasari\thanksref{headdress} }
and 
\author[yy]{Yoshio Yamaguchi}

\address[hk]{Department of Physics, 
School of Science, Tokai University, 
Hiratuka, Kanagawa 259-1292}
\address[yy]{c/o RIKEN,  Wako, Saitama, 351-0198}

\thanks[headdress]{E-mail address: kasari@keyaki.cc.u-tokai.ac.jp}

\begin{abstract}
Contrary to the conventional belief, it was shown that the Breit 
equation has the eigenvalues for bound states of two oppositely charged 
Dirac particles interacting through the (static) Coulomb potential. 
All eigenvalues reduced to those of the Sch\"odinger case in the 
non-relativistic limit. 
\end{abstract}

\end{frontmatter}

Consider the bound states of two oppositely (single) charged Dirac particles 
intercting through the (static) Coulomb potential.
They can be described by the Breit equation\cite{rf:1,rf:2,rf:3}
\begin{eqnarray}
\left [ (\vec{\alpha}_{1} \frac{1}{i}\vec{\nabla} + 
\beta_{1}m)_{\alpha \alpha^{\prime}} 
\delta_{\beta \beta^{\prime}} + \delta_{\alpha \alpha^{\prime}} 
(-\vec{\alpha}_{2}\frac{1}{i}\vec{\nabla} + \beta_{2}M)_{\beta \beta^{\prime}} 
- \frac{\alpha}{r} \delta_{\alpha \alpha^{\prime}} 
\delta_{\beta \beta^{\prime}} \right ] \Psi_{\alpha^{\prime} \beta^{\prime}} 
\nonumber \\
= E \Psi_{\alpha \beta},
\end{eqnarray}
where we take the specific Lorentz frame (the CM system) and the gauge 
(Coulomb gauge), $\vec{\alpha_{j}}, \beta_{j} (j=1,2)$ are  Dirac matrices 
of the particle $j$, 
while $m$ and $M$ are their masses.
$\Psi_{\alpha \beta}$ is the $4\times4$ Dirac spinor wave function 
$(\alpha, \beta = 1, \cdots, 4)$. 
The total energy of the system (in the CM system) is writen as 
\begin{eqnarray}
E = \sqrt{m^2 - q^2} + \sqrt{M^2 - q^2}
\end{eqnarray}
where a discrete values of $q$ coressponds to a bound state.

It is convenient to introduce the dimensionless quantities
\begin{eqnarray}
\left.
\begin{array}{ccl}
 \rho & = & 2 q r,  \\ [0.3cm]
 y & = & \frac{E}{2\alpha q} = 
  \frac{\sqrt{m^2 - q^2} + \sqrt{M^2 - q^2}}{2\alpha q}, \\ [0.3cm]
\lambda & = & \frac{M + m + E}{2\alpha q}, \\ [0.3cm]
\nu & = & \frac{M + m - E}{2\alpha q},
\end{array}
\right\}
\label{eq:3}
\end{eqnarray} 
\begin{eqnarray}
\alpha \; = \; e^2 / \hbar c \; 
\lower.8ex\hbox{$\stackrel{\doteq}{ .}$} \; 1/137.
\label{eq:4}
\end{eqnarray}

First, we discuss the simplest case\cite{rf:4}: 
the singlet state and equal mass case $M = m$. 
Let $F(r)$ and $K(r)$ be the radial wave functions of the large-large 
$(\alpha, \beta = 1, 2)$ and small-small $(\alpha, \beta = 3, 4)$ 
components of $\Psi_{\alpha\beta}$. 
Then we find from Eq.(1)
\begin{eqnarray}
K & = & \frac{1 - \nu \rho}{1 + \lambda \rho} F, \\
F + K & = & \tilde{h}(\rho) e^{-\rho/2}
\end{eqnarray}
and $\tilde{h}(\rho)$ should satisfy
\begin{eqnarray}
\frac{d^2 \tilde{h}}{d \rho^2} &+& 
\frac{d \tilde{h}}{d \rho}\left\{-1 + \frac{2}{\rho} + 
                        \frac{1}{\rho(\rho + y \rho)} \right\} \nonumber \\
&+& \tilde{h} \left\{ \left( \frac{\alpha^2 y}{2} -1 \right) \frac{1}{\rho} 
  + \frac{\alpha^2}{4} \frac{1}{\rho^2} - \frac{l(l + 1)}{\rho^2} 
  - \frac{1}{ 2 \rho (1 + y \rho)} 
   \right\}  = 0,
\label{eq:7}   
\end{eqnarray}   
where $l$ is the orbital angular momentum, $l = 0, 1, 2, \cdots.$

The non-relativistic limit of Eq.(7) means $\alpha^2 \rightarrow 0$ 
there, while keeping $\alpha^2 y \rightarrow $ const. 
Ignoring terms in $\alpha^2$ in Eq.(7) and noticing $1/y$ is of the 
order of $\alpha^2$, Eq.(7) reduces to 
\begin{eqnarray}
\frac{d^2 \tilde{h}}{d \rho^2} + 
\frac{d \tilde{h}}{d \rho}\left\{-1 + \frac{2}{\rho}  \right\} 
+ \tilde{h} \left\{ \left( \frac{\alpha^2 y}{2} -1 \right) \frac{1}{\rho} 
   - \frac{l(l + 1)}{\rho^2} 
    \right\}  = 0,
\label{eq:8}   
\end{eqnarray} 
which is precisely the Schr\"{o}dinger case. 
Therefore Eq.(7) is the proper relativistic extension in the singlet case.

Let us take $l=0$. Near $\rho = 0 \ \tilde{h}(\rho)$ can be expressed as 
\begin{eqnarray}
\tilde{h}(\rho) = \rho^{s} h(\rho),
\label{eq:9}   
\end{eqnarray}
where
$s = -1 + \sqrt{1 - \frac{\alpha^2}{4}}$.
Another solution with $s = -1 - \sqrt{1 - \frac{\alpha^2}{4}}$ is 
not acceptable as the wave function. 
$h(\rho)$ obeys the equation 
\begin{eqnarray}
h'' &+& 
\left\{ -1 + \frac{2+2s}{\rho} + \frac{1}{\rho(\rho + y \rho)} \right\} h' 
\nonumber \\
 &+& \left\{ \left( \frac{\alpha^2 y}{2} -1 - s \right) \frac{1}{\rho} 
  - \left( \frac{1}{2} + s y \right) \frac{1}{\rho (1 + y \rho)} 
   \right\} h = 0.
\label{eq:10}   
\end{eqnarray}   

$h(\rho)$ can be expressed in the Taylor series near $\rho = 0$ 
$(|\rho| < \frac{1}{y})$.
\begin{eqnarray}
h(\rho) = \sum^{\infty}_{n=0} h_n \rho^n,
\label{eq:11}   
\end{eqnarray}
which is very much different from the Sch\"{o}dinger case 
(Laguerre polynomials) in the small domain $0 < \rho < \frac{1}{y}$.
Perhaps this led peoples in old days the belief that the Breit equation 
is too singular and should not be solved it directly 
(but use the perturbation to deal with it to obtain reasonable 
eigenvalues of bound states\cite{rf:2,rf:3}).
We should notice that $\tilde{h}(\rho)$ is reasonably close to 
the corresponding non-relativistic solution at $\rho \gg 1/y$ 
(see Figs. 3 and 6 shown below).

We observe that $\rho = \infty$ is not the regular singular point in Eq.(10). 
There are two independent solutions near $\rho = \infty$ of the Eq.(10):
\begin{eqnarray}
h_1(\rho) = \rho^{\beta} \left\{ 1 + \frac{c_1}{\rho} + \frac{c_2}{\rho^2} 
+ \cdots \right\}
\label{eq:12}   
\end{eqnarray}
and
\begin{eqnarray}
h_2(\rho) = e^{\rho} \rho^{\beta^{'}} 
\left\{ 1 + \frac{c^{'}_1}{\rho} + \frac{c^{'}_2}{\rho^2} 
+ \cdots \right\}
\label{eq:13}   
\end{eqnarray}
where $\beta$ and $\beta^{'}$ are undetermined from Eq.(10) 
near $\rho = \infty$.
On the other hand we may analytically continue the solution 
$h(\rho)$ in Eq.(11) valid near $\rho = 0$ to $\rho \rightarrow \infty$. 
Then we should find 
\begin{eqnarray*}
h(\rho) &=& c_1 h_1(\rho) + c_2 h_2(\rho)  \ \ \ \ \mbox{near} 
                                           \ \ \ \ \rho = \infty,
\end{eqnarray*}
where $c_1, c_2, \beta$ and $\beta^{'}$ depend on the parameters in Eq.(10), 
namely $\alpha^2$ (or $s$) and $y$.
Quantum Mechanics demands $c_2$ should be zero, which fixes allowed values 
of $y$, and hence those of $E$, the energy eigenvelues. 
Equivalently we can impose the condition: 
After analitic continuation the solution $h(\rho)$, Eq.(11), near $\rho = 0$ 
should smoothly join the solution $h_1(\rho)$, Eq.(12), near $\rho = \infty$. 
This condition selects the specific discrete values of $\beta$ 
(which is a function of $\alpha^2$ and $y$), 
giving specific values of $y$ 
(which must agree with the values of $y$ derived from $c_2 = 0$).

The Eq.(10) is known as Heun's equation, we owe to Professor Th. A. Rijken
 and his son Pieter to informed us of this information. 
 However we have not yet succeeded to find the analytic form of $\beta$ 
 or $c_2$.
 
Nevertheless the perturbative consideration of Eq.(10) shows that for small 
$\alpha$
\begin{eqnarray*}
\beta  & = & n - 1 - s + O(\alpha^4), \\ 
n & = & 1, 2, \cdots, 
\end{eqnarray*}
where perturbation is with respect to $s$ or $\alpha^2$, and note that 
\begin{eqnarray*}
\frac{1}{y} & \sim & O(\alpha^2), \\
s & \sim & - \frac{1}{8} \alpha^2.
\end{eqnarray*}

To be more precise we have tried to obtain an accurate value of $\beta$ 
by computer for the lowest two state ($n = 1$ and $2$) in the singlet case. 
Changing a value of $\beta$, we tried to get smooth joining of two solutions 
$h(\rho)$ starting from $\rho = 0$ and $h_1(\rho)$ starting 
from $\rho = \infty$ as good as possible. 
Table I shows our results, where the minimum value of 
\begin{eqnarray}
\triangle = \left | \frac{1}{h} \frac{d h}{d \rho} - 
\frac{1}{h_1} \frac{d h_1}{d \rho} \right | = 0 
\label{eq:14}   
\end{eqnarray}
is found at $\rho = \rho_1$. 
\setcounter{footnote}{0}
\begin{table}
 \caption{ 
 The Breit equation (singlet, ground state ($n=1$), equal mass $M=m$ case) 
 is solved by computer. 
 $\alpha$ is the fine structure constant.
 However, we have computed $\beta$ for various values of $\alpha$.
 $s$ is given by Eq.~(9).
 $\beta$ is determined by computer calculation.
  $\Delta$ is minimized discrepancy of the logrithmic derivatives 
 between the solution from 
 $\rho = 0$ and the solution from $\rho \rightarrow \infty$, 
 is found at $\rho_1$, see Eq.~(14). 
 }
  \label{tabel:1}
 \begin{center}
  \begin{tabular}{cclcl} \hline 
  $\alpha^2$ & $-s$ & \ \ \ \ \ \ \ \ \ \ \ $\beta$ & $\rho_1$ 
             & \ \ \ \ \ $\Delta$ \\ \hline 
  3 & 0.5& 0.1712475 & 1.7 & $0\  \sim \ O(10^{-11})$ \\ 
  2 & 0.2928932 &  0.1165384 & 1.9 & $0\  \sim \ O(10^{-11})$ \\  
  1 & 0.1339746 &  0.06732758 & 2.0 & $0\  \sim \ O(10^{-11})$ \\  
  $\frac{1}{2}$ & 0.06458565 & 0.03945101 & 2.0 & $0\  \sim \ O(10^{-11})$ \\ 
                                                                     
  $\frac{1}{10}$ & 0.01257912 & 0.01034584 & 2.0 & $0\ \sim \ O(10^{-12})$ \\ 
                                                                    
  $\frac{1}{100}$ & 0.001250782 & 0.001203460 & 2.0 & 
  $0\  \sim \ O(10^{-13})$ \\   
 $\frac{1}{1000}$ & 0.0001250078 & 0.0001242537 & 2.0 & 
 $0\  \sim \ O(10^{-14})$ \\   
 $\frac{1}{10000}$ & 0.00001250008 & 0.00001248938 & 2.0 & 
 $0\  \sim \ O(10^{-14})$  \\  
 $\frac{1}{(136)^2}$ & $6.758241 \times 10^{-6}$ & 
     $6.754162 \times 10^{-6}$ & 2.0 & $0\  \sim \ O(10^{-14})$ \\    
 $\frac{1}{(137)^2}$ & $6.659940 \times 10^{-6}$ & 
     $6.655932 \times 10^{-6}$ & 2.0 & $0\  \sim \ O(10^{-14})$ \\  
 $\frac{1}{(137.035999765)^2}$ & $6.656441  \times 10^{-6}$ & 
     $6.652436  \times 10^{-6}$  & 2.0 & $0\  \sim \ O(10^{-14})$ \\   
 $\frac{1}{(138)^2}$ & $6.563769 \times 10^{-6}$ & 
     $6.559828 \times 10^{-6}$ & 2.0 & $0\  \sim \ O(10^{-14})$ \\  
 $\frac{1}{20000}$ & $6.250020 \times 10^{-6}$ & 
     $6.246287 \times 10^{-6}$ & 2.0 & $0\  \sim \ O(10^{-14})$ \\ \hline 
   \end{tabular}
 \end{center}
\end{table}
We see there 
\begin{eqnarray*}
\beta \ \lower.8ex\hbox{$\stackrel{\doteq}{ .}$} - s
\end{eqnarray*}
is reasonably good when $\alpha \ll 1$.
We found 
\begin{eqnarray*}
\beta \ &\lower.8ex\hbox{$\stackrel{\doteq}{ .}$}& -s \ \{ 1 + 90.5 \ s \} 
\ \ \ \hbox{for} \  n = 1
\end{eqnarray*}
or
\begin{eqnarray*}
\beta \ &\lower.8ex\hbox{$\stackrel{\doteq}{ .}$} & -s \ 
\{ 1 - 11.3 \ \alpha^2 \}, \ \ (n= 1)
\end{eqnarray*}
near $\alpha= 1/137.036$.
The relations between $\beta$ versus $s$ or $\alpha^2$ are depicted in 
Figs.1, 2, 4 and 5
 (for n = 1 and 2, respectively).

\begin{figure}
 \parbox{6.7cm}{
   \epsfxsize= 6.6cm 
 \epsfbox{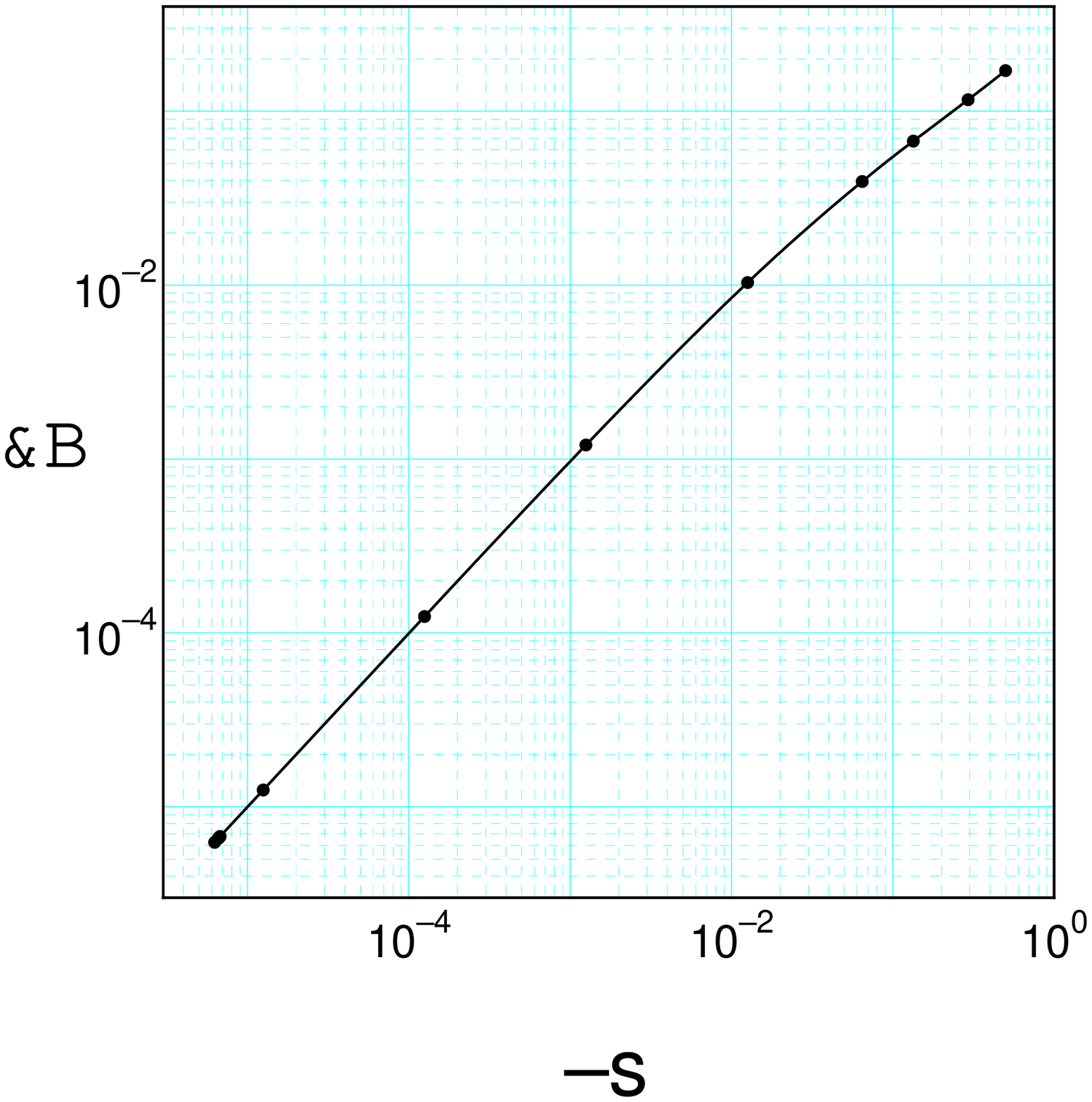}
 \caption{The relatons between $\beta$ versus $-s$ 
 for the singlet ground state ($n=1$).}
 \label{fig:1} 
                   }
 \hspace{8mm}
\parbox{6.7cm}{  
  \epsfxsize= 6.6cm 
  \epsfbox{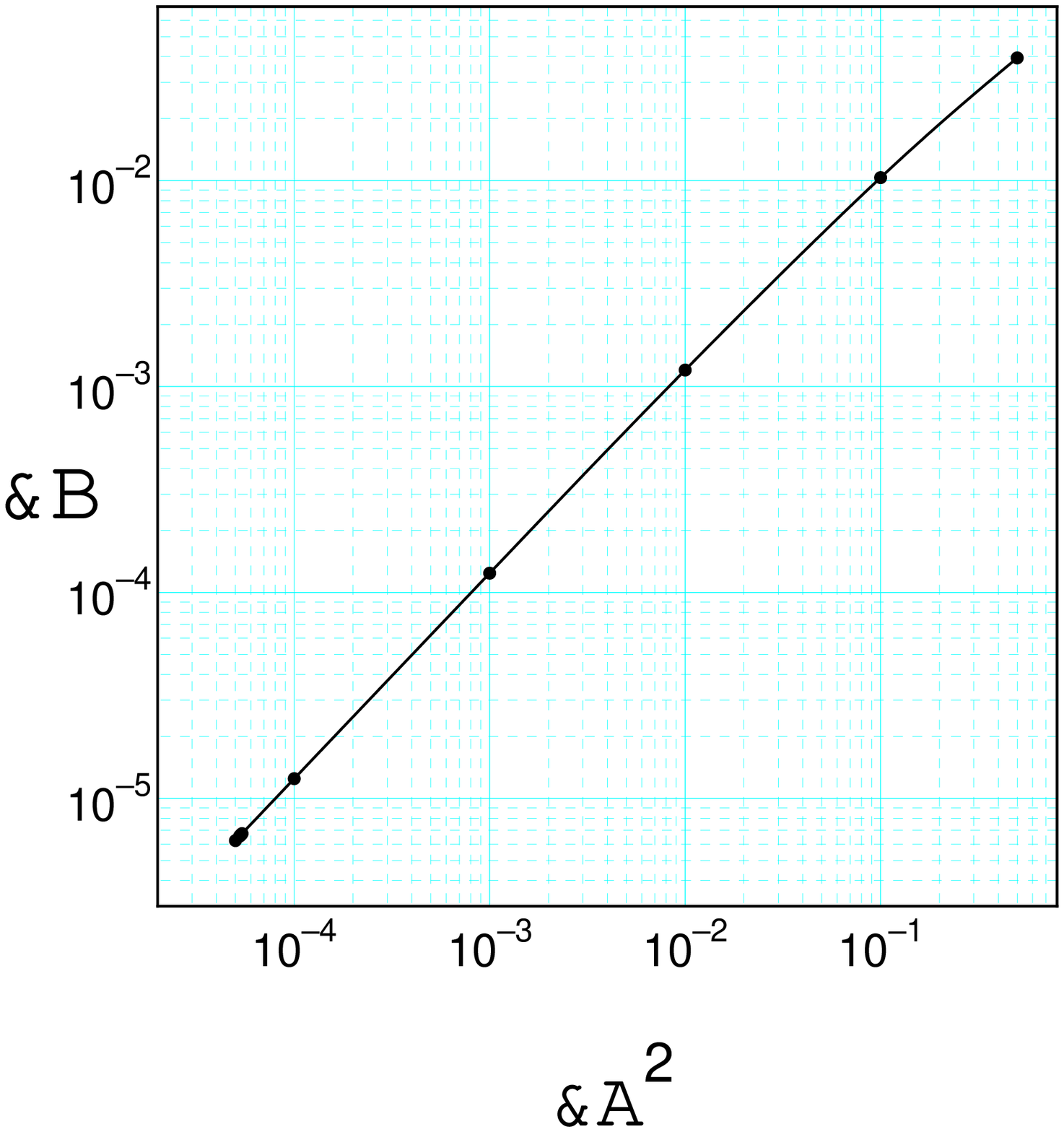}
  \caption{
  The relatons between $\beta$ versus $\alpha^2$ 
  for the singlet ground state ($n=1$).
 }
 \label{fig:2}
     }
\end{figure}

We also illustrate the best fit (smallest $\triangle$) of the 
``radial wave function" $h(\rho) = (F + K) \ e^{\rho/2}$ 
for the ground state of ${}^1S$ in Fig.3

\begin{figure}
 \centerline{\epsfbox{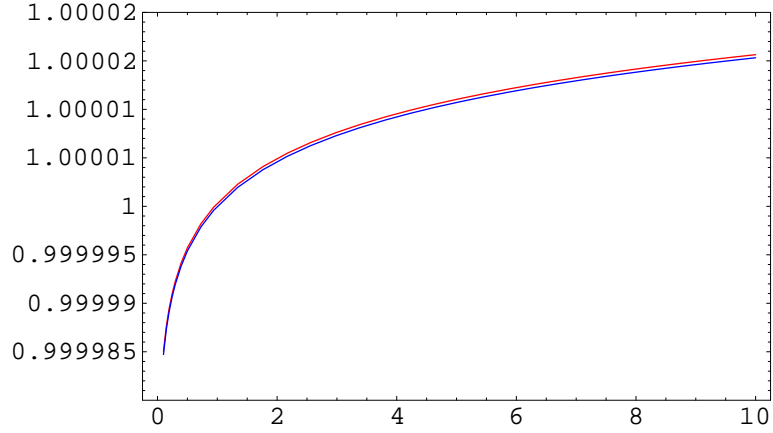}}
 \caption{
 The ground state ($n=1$) wave function $h$ and $h_1$ for 
 $\alpha^2 = \frac{1}{(137.035999765)^2}$ and 
 $\beta = - s + (\delta \beta) = 6.65244 \times 10^{-6}$.
 The horizontal axis denotes the radial parameter $\rho$.
 The vertical axis denotes values of the wave function 
 $h (\rho)$ and $h_1(\rho)$.
 The red line denotes $h(\rho)$ from $\rho = 0$. 
 The blue one denotes $h_1(\rho)$ from $\rho = \infty$.
 $h (\rho)$ and $h_1 (\rho)$ are very close 
 to the non-relativistic solution 1.
   }
 \label{fig:3}
\end{figure} 
Once  $\beta$ is known the eigenvalues $E$ can easily be found from Eq.(10)
\begin{eqnarray*}
- \beta + \left\{ \mbox{coefficient of} \ \frac{1}{\rho} \ \mbox{in Eq.(10)} 
 \right\} = 0,
\end{eqnarray*}
or 
\begin{eqnarray*}
\frac{\alpha^2 y}{2} -1 -s &=& \beta \equiv n -1 -s + (\delta \beta) \\
 &=& n -1 -s + O(\alpha^4)
\end{eqnarray*}
where
\begin{eqnarray*}
y = \frac{\sqrt{m^2 - q^2}}{\alpha q} \ \ \hbox{and} \ \ n = 1, 2, 3, \cdots. 
\end{eqnarray*}
The binding energy is given by
\begin{eqnarray*}
B &=& 2 m - E = 
 2m \left[ 1 - \frac{1}{\sqrt{1 + \frac{\alpha^2}{2(n + \delta \beta)}}} 
    \right], \\ 
n &=& 1, 2, \cdots, \\ 
\delta \beta &=& O(\alpha^4).
\end{eqnarray*}
This result agrees with existing perturbative results 
according to M. Nio\cite{rf:5}.

 Next we also calculate the value of $\beta$ for the excited state 
(n = 2) in the singlet case.
The results are shown in Table II, Figs.4, 5 and 6.

\setcounter{footnote}{0}
\begin{table}
 \caption{ 
 The Breit equation (singlet, excited state ($n=2$), equal mass $M=m$ case) 
 is solved by computer. 
 $\alpha$ is the fine structure constant.
 However, we have computed $\beta$ for various values of $\alpha$.
 $s$ is given by Eq.~(9).
 $\beta$ is determined by computer calculation.
  $\Delta$ is minimized discrepancy of the logrithmic derivatives 
 between the solution from 
 $\rho = 0$ and the solution from $\rho \rightarrow \infty$, 
 is found at $\rho_1$, see Eq.~(14). 
 }
  \label{tabel:1}
 \begin{center}
  \begin{tabular}{cclcl} \hline  
  $\alpha^2$ & $-s$ & \ \ \ \ $\beta -1$ & $\rho_1$ 
             & \ \ \ \ \ $\Delta$ \\ \hline 
  3 & 0.5& 0.1237607 & 4.4 & $0\  \sim \ O(10^{-10})$ \\ 
  2 & 0.2928932 &  0.09529568 & 4.9 & $0\  \sim \ O(10^{-10})$ \\  
  1 & 0.1339746 &  0.06081300 & 5.1 & $0\  \sim \ O(10^{-10})$ \\ 
  $\frac{1}{2}$ & 0.06458565 & 0.03739075 & 5.2 & $0\  \sim \ O(10^{-10})$ \\ 
                                                                     
  $\frac{1}{10}$ & 0.01257912 & 0.01022230 & 5.2 & $0\ \sim \ O(10^{-10})$ \\ 
                                                                     
  $\frac{1}{100}$ & 0.001250782 & 0.001201915 & 5.2 & 
  $0\  \sim \ O(10^{-10})$ \\  
 $\frac{1}{1000}$ & 0.0001250078 & 0.0001242386 & 5.2 & 
 $0\  \sim \ O(10^{-10})$ \\   
 $\frac{1}{10000}$ & 0.00001250008 & $1.250836 \times 10^{-5}$ & 5.2 & 
 $0\  \sim \ O(10^{-10})$  \\ 
 $\frac{1}{(136)^2}$ & $6.758241 \times 10^{-6}$ & 
     $6.792039 \times 10^{-6}$ & 5.2 & $0\  \sim \ O(10^{-10})$ \\    
 $\frac{1}{(137)^2}$ & $6.659940 \times 10^{-6}$ & 
     $6.694315 \times 10^{-6}$ & 5.2 & $0\  \sim \ O(10^{-10})$ \\  
 $\frac{1}{(137.035999765)^2}$ & $6.656441  \times 10^{-6}$ & 
     $6.691002 \times 10^{-6}$  & 5.2 & $0\  \sim \ O(10^{-10})$ \\   
 $\frac{1}{(138)^2}$ & $6.563769 \times 10^{-6}$ & 
     $6.599096 \times 10^{-6}$ & 5.2 & $0\  \sim \ O(10^{-10})$ \\  
 $\frac{1}{20000}$ & $6.250020 \times 10^{-6}$ & 
     $6.287715 \times 10^{-6}$ & 5.2 & $0\  \sim \ O(10^{-10})$ \\ \hline 
   \end{tabular}
 \end{center}
\end{table}
  
\begin{figure}
\parbox{6.7cm}{  
   \epsfxsize= 6.6cm 
 \epsfbox{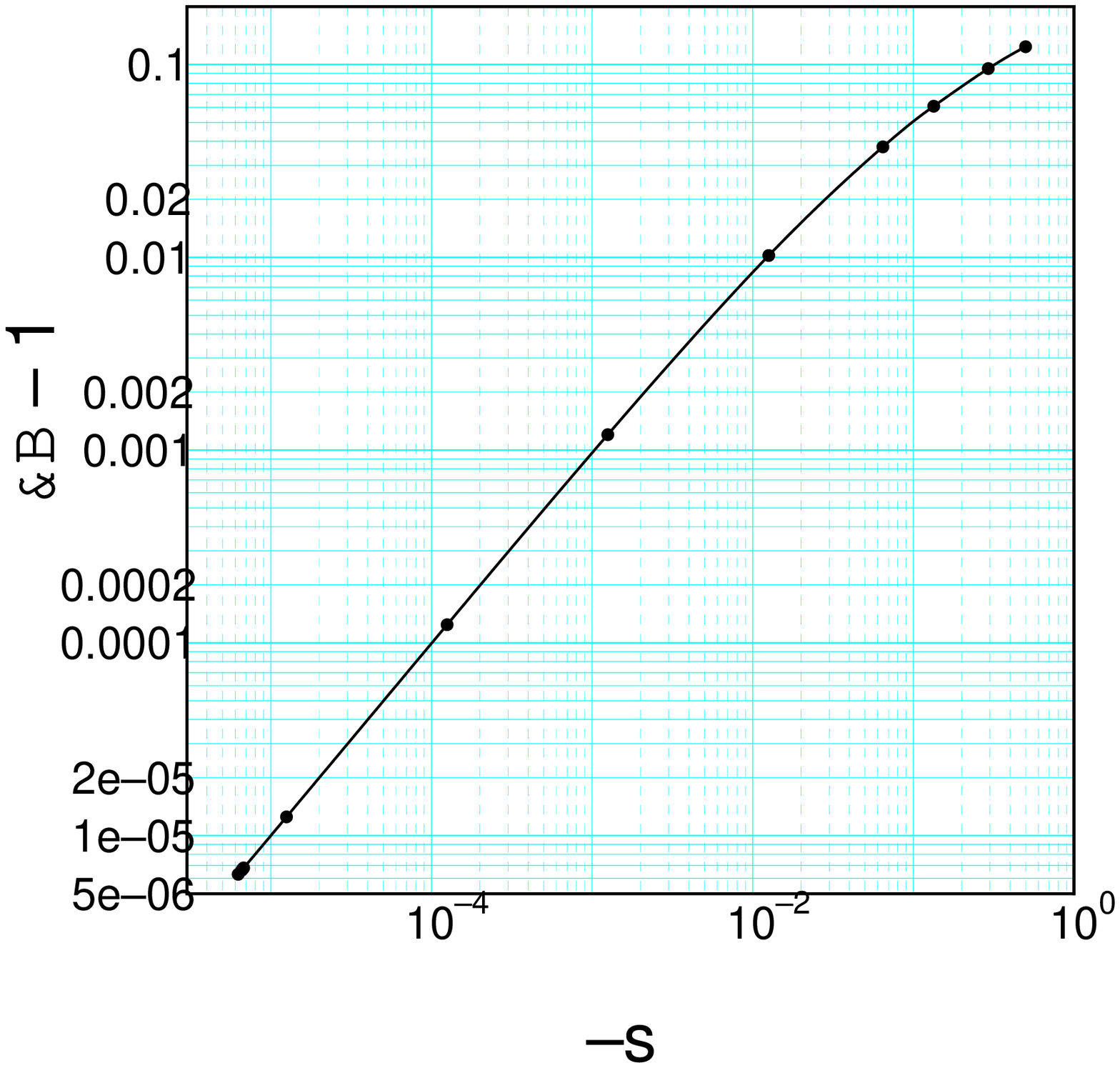}
 \caption{
 The relatons between $\beta - 1$ versus $-s$ 
 for the singlet excited state ($n=2$).
 }
 \label{fig:4}
    }
 \hspace{8mm}
 \parbox{6.7cm}{  
  \epsfxsize= 6.6cm 
  \epsfbox{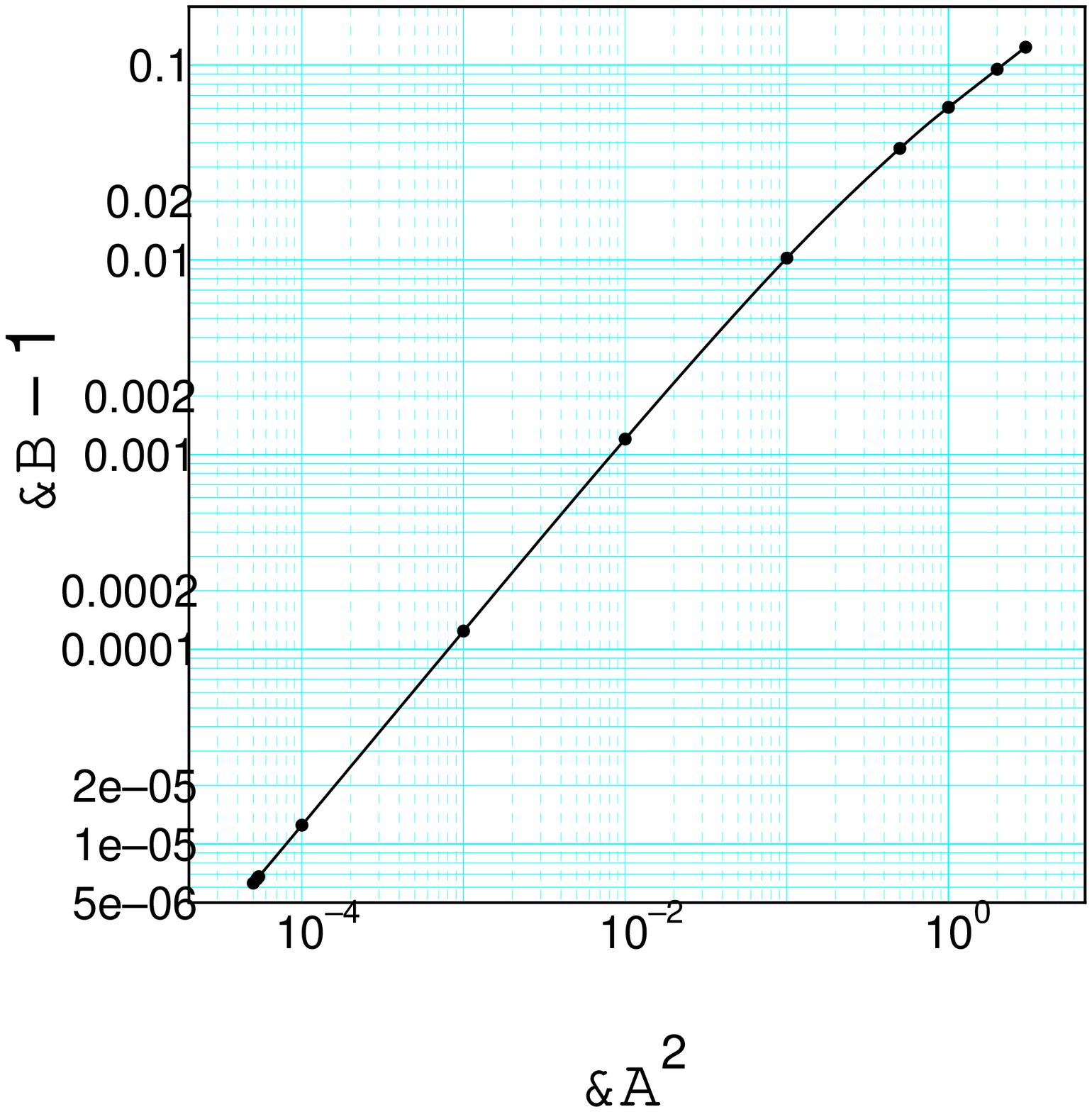}
  \caption{
  The relatons between $\beta - 1$ versus $\alpha^2$ 
  for the singlet excited state ($n=2$).
 }
 \label{fig:5}
     }
\end{figure}
\begin{figure}
 \centerline{\epsfbox{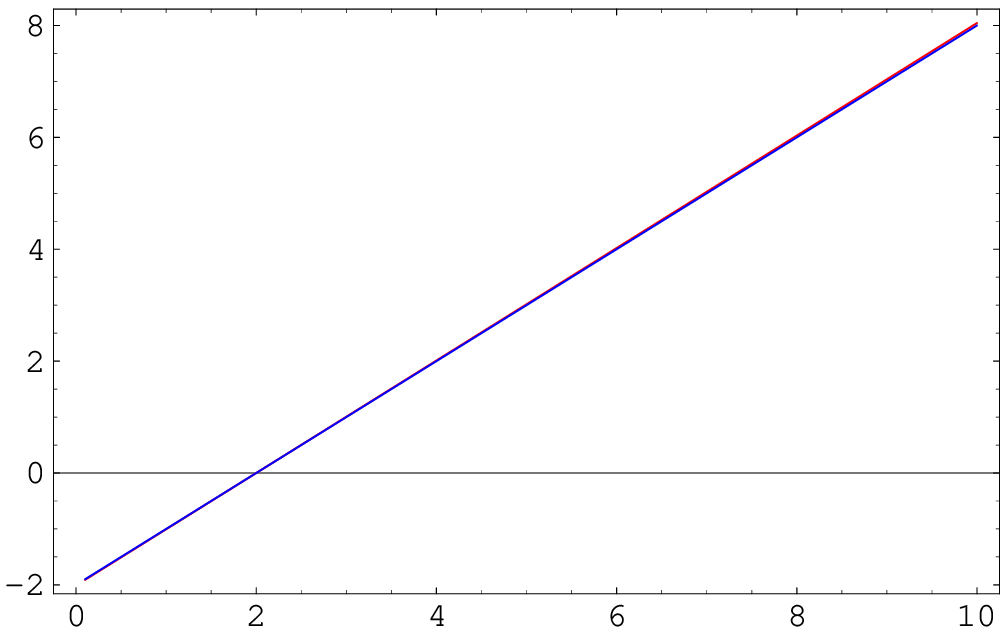}}
 \caption{
 The excited state ($n=2$) wave function $h$ and $h_1$ for 
 $\alpha^2 = \frac{1}{(137.035999765)^2}$ and 
 $\beta = - s + (\delta \beta) = 1.00001$.
 The horizontal axis denotes the radial parameter $\rho$.
 The vertical axis denotes velues of the wave function $h (\rho)$.
 The red line denotes $h(\rho)$ from $\rho = 0$. 
 The blue one denotes $h_1(\rho)$ from $\rho = \infty$.
 $h (\rho)$ and $h_1 (\rho)$ are very close 
 to the non-relativistic solution $1 - \frac{\rho}{2}$
   }
 \label{fig:6}
\end{figure} 

Up to now we have discussed the simplest case, ${}^1S$ and $M=m$.
The above consideration can be applied in general cases 
(singlet and triplet cases and also $M \neq m$) 
of the Breit equation where not only the Coulomb potential 
but also the one transverse photon exchange contribution are 
taken into account
\cite{rf:4}. 
Now we can conclude that the Breit equation gives eigen functions and 
eigenvalues for the Coulombian bound states, which are proper relativistic 
generalization of the (non-relativistic) Schr\"{o}dinger case.

\end{document}